\documentclass[a4paper,11pt]{article}
\usepackage{jinstpub} 

\usepackage{booktabs}
\usepackage{subfigure}
\usepackage{makecell}

\title{\boldmath Study on the Impact of Radioactive Background on the Dark Count Rate of 20‑inch MCP‑PMTs}

\author[a]{Zeyuan Feng,}
\author[b,c]{Zhaoyuan Peng,}
\author[b,c]{Haojie Dong,}
\author[b]{Yanfeng Li,}
\author[b]{Songyi Li,}
\author[a]{Xinzhou Guo,}
\author[b]{Wan Xie,}
\author[b,1]{Zhonghua Qin\note{Corresponding author.}}
\author[a,2]{and Weiping Liu\note{Corresponding author.}}

\affiliation[a]{Department of Physics, Southern University of Science and Technology, Shenzhen, China}
\affiliation[b]{Institute of High Energy Physics, Beijing 100049, China}
\affiliation[c]{University of Chinese Academy of Sciences, Beijing 100049, China}

\emailAdd{qinzh@ihep.ac.cn}
\emailAdd{liuwp@sustech.edu.cn}

\abstract{This study systematically investigates the impact of natural radioactive background on the dark count rate (DCR) of 20-inch microchannel plate photomultiplier tubes (MCP-PMTs). Variations in PMTs' DCR under different radiation conditions were examined via underground tests, lead shielding experiments, and irradiation with \(^{55}\mathrm{Fe}\), \(^{60}\mathrm{Co}\), and \(^{90}\mathrm{Sr}\) sources. The experimental results indicate that natural radioactivity from surrounding air and rock in the underground environment results in a significantly higher DCR compared to laboratory conditions. Further, lead shielding experiments confirmed that effective shielding can markedly reduce the interference from environmental background radiation. Notably, $\beta$ particles from the \(^{90}\mathrm{Sr}\) source increased the DCR by approximately 14 kHz, whereas the effects of $\mathrm{X}$-rays from \(^{55}\mathrm{Fe}\) and $\gamma$-rays from \(^{60}\mathrm{Co}\) were comparatively minor. In addition, Geant4 simulations provided quantitative analysis of Cherenkov radiation induced by $\beta$ particles, with the simulation results closely matching the experimental data and confirming $\beta$ particles as the primary contributor to the DCR increase. These findings offer both theoretical and experimental evidence for a deeper understanding of the influence of radioactive background on 20-inch MCP-PMTs' performance in underground experiments and hold significant implications for improving the energy resolution of large-scale neutrino detection systems}


\begin{document}
\maketitle
\flushbottom

\section{Introduction}
The photomultiplier tube (PMT) is a device that can convert weak light signals into measurable electrical signals. It is widely used in various fields such as particle physics\cite{bellini2014neutrino}, astronomical observations, and life sciences. Particularly in particle physics experiments, PMTs play a crucial role as key photon detection devices. In neutrino experiments, such as  the SNO experiment\cite{ahmad2002SNO}, the Super-Kamiokande experiment\cite{fukuda1998superK-1,wendell2010SuperK-3}, the IceCube experiment\cite{halzen2010IceCube} and the Daya Bay experiment\cite{cao2016overviewDayaBay}, have all extensively employed large numbers of PMTs as their primary photon detection devices.

PMTs can be classified into Dynode-type and Microchannel plate-type\cite{anashin1995photomultipliers}. The 20-inch Microchannel plate PMT (MCP-PMT) is a newly developed large photocathode area PMT\cite{WANG2012113}. In both the Jiangmen Underground Neutrino Observatory (JUNO)\cite{juno2022juno} and the Large High Altitude Air Shower Observatory (LHAASO)\cite{lhaaso2010future}, a large number of 20-inch MCP-PMTs have been employed.

The performance of PMTs is essential for achieving the physics goals of particle physics experiments described above. Several parameters are commonly used to characterize PMTs performance, including quantum efficiency, collection efficiency, and dark count rate (DCR)\cite{abusleme2022mass}. The sources of dark noise in PMTs are diverse, including thermionic emission from the photocathode and multiplication stages\cite{wilson2023characterisation}, ionization of residual gases within the vacuum enclosure. Furthermore,external photons can also elevate the PMTs' DCR. Luminescence generated by the glass envelope and structural materials, Scintillation photon emitted by the MCP\cite{ren2022contribution} and Cherenkov photon generated by comic ray\cite{zhang2022study}. These photon-induced counts are included in the DCR, thereby further raising its measured value. Consequently, any evaluation of the DCR must explicitly account for the contribution of externally induced photon signals.

As one of the critical parameters affecting PMTs' performance, DCR is a central focus in the development of PMTs. This is because DCR influences the level of experimental background noise, and particularly in low-background neutrino experiments, it can interfere with genuine signal counting and be misidentified as photons produced by particle events, thereby degrading the energy resolution of the detector. In certain experiments, the impact of the DCR on energy resolution can reach as high as approximately $30\%$\cite{abusleme2021calibration,abusleme2025prediction}. As a novel large-area photon sensor, the 20-inch MCP-PMT requires a thorough understanding of the sources and underlying mechanisms of its dark noise, which is essential for effectively reducing its noise level and enhancing overall performance.

The influence of natural radioactivity on the DCR of MCP-PMTs is also non-negligible. The natural radioactive background primarily originates from radioactive substances present in the surrounding rocks and air, which emit $\alpha$, $\gamma$, and $\beta$ radiation during their decay processes. To systematically investigate the mechanisms by which natural radioactivity affects the DCR, this study has designed a series of experiments. First, underground experiment were conducted to evaluate the influence of background radiation. Subsequently, the MCP-PMT was tested in the lead shielding environment. In addition, irradiation experiments with radioactive sources were performed to observe variations in DCR under controlled exposure. Furthermore, Geant4 simulations were employed to model the interaction of radiation with the MCP-PMTs, enabling a detailed analysis of the mechanisms by which different types of radiation influence the DCR.

\section{Experimental Study on the Impact of Natural Radioactive Background}
\subsection{Underground Experiment}\label{section2.1}
The level of natural radioactive radiation in underground environments is typically higher than that at the surface. To investigate whether natural radioactivity affects the DCR of PMTs, we conducted a PMT underground experiment in the experimental hall of the JUNO detector. 

\begin{figure}
    \centering
    \includegraphics[width=0.7\linewidth]{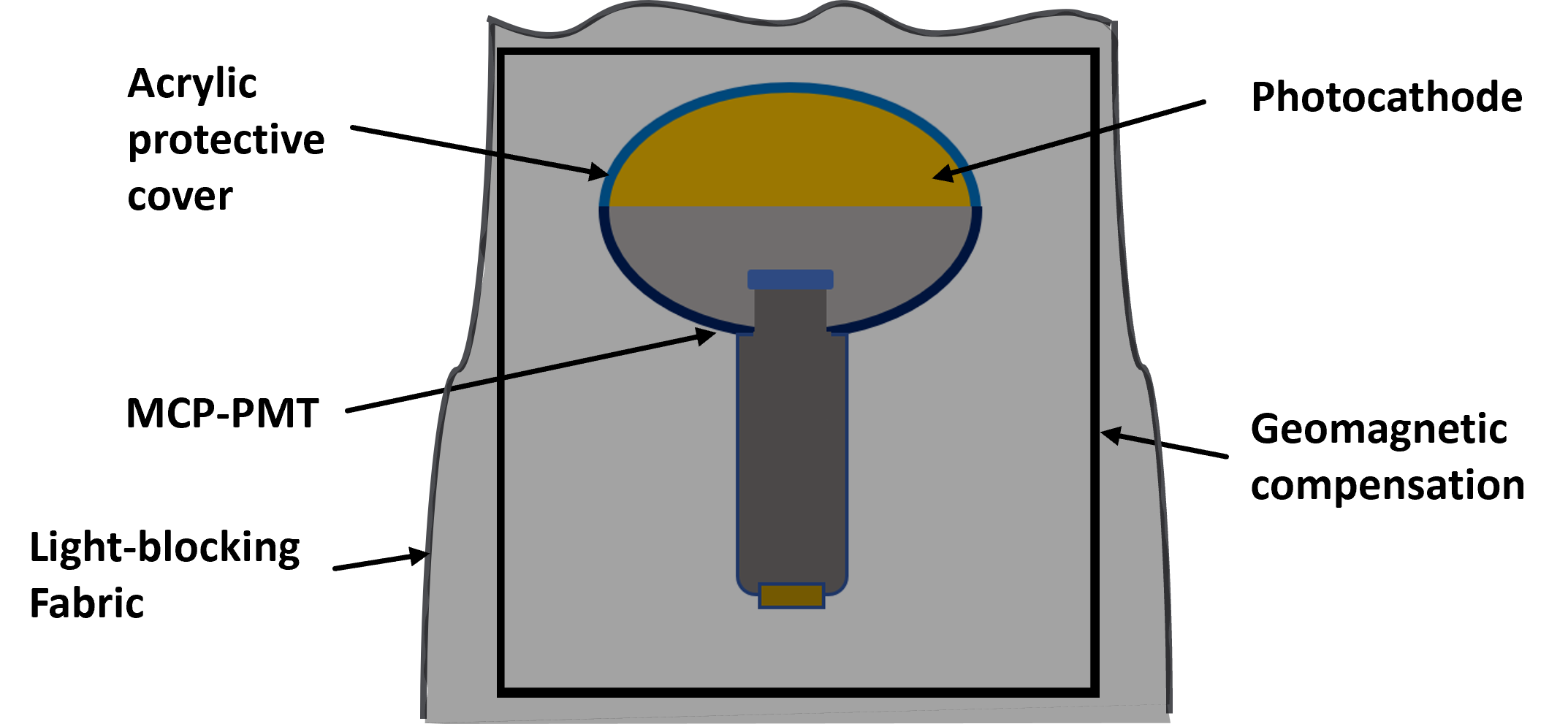}
    \caption{Underground Experiment Setting up}
    \label{fig:undergroundExpSetup}
\end{figure}

We placed PMTs near the central detector of JUNO to simulate the operational conditions that PMTs would experience in a realistic underground environment as indicated by figure~\ref{fig:undergroundExpSetup}. In the experiment, we used a light-tight box with geomagnetic field compensation materials\cite{nanocrystalline} and covered it with black cloth to create a completely dark environment, ensuring reliable experiment conditions. A total of five PMTs were tested, with the results shown in table~\ref{tab:PMT_underground_exp}, the difference is defined as the result measured in the surface laboratory testing minus that measured in the underground experiment, the last row of the table presents the average values of the experimental results. With regard to the systematic uncertainty, the high voltage applied during the tests was not recalibrated to the exact operating value that yields a gain of $10^7$ after encapsulation. This contributes an uncertainty of less than $1~\mathrm{kHz}$ to the reported DCR. The DCR was measured with a self-trigger algorithm implemented directly in the FPGA of the readout electronics. Any signal exceeding a predefined threshold was registered as a valid dark count. This threshold, corresponding to approximately 0.25 photoelectrons, was established through a preliminary threshold scanning procedure\cite{peng2025test}. The method reliably distinguishes genuine photoelectron pulses from electronic noise, thereby ensuring an accurate determination of the PMT dark noise level. For each run, the measured value was taken as the PMT’s steady-state DCR only after the rate had levelled off and all after-exposure transients had vanished—a period we refer to as the ‘cooling down process’. During acquisition the DCR was sampled continuously at $1~\mathrm{s}$ intervals. To secure sufficient statistics, each measurement run typically lasted approximately 3 hours. When necessary, the acquisition was further extended until the stability criterion was satisfied.

\begin{table}[htbp]
  \centering
  \caption{Comparison of PMTs' DCR in underground and laboratory environments}
  \begin{tabular}{lccc}
    \toprule
PMT Serial Number & \makecell{Underground\\experiment result (kHz)} & \makecell{Surface laboratory\\testing result (kHz)} & \makecell{Difference\\(kHz)} \\
    \midrule
    PA1711-1570 & 34.30 & 27.28 & 7.02   \\
    PA1712-1149 & 52.93 & 21.29 & 31.64  \\
    PA1704-679  & 56.63 & 45.10 & 11.53  \\
    PA1809-1283 & 38.05 & 18.70 & 19.35  \\
    PA2007-1183 & 61.95 & 39.61 & 22.34  \\
    The average & 48.78 & 30.40 & 18.38  \\
    \bottomrule
  \end{tabular}
  \label{tab:PMT_underground_exp}
\end{table}
The results indicate a significant increase in the DCR of PMTs in the underground environment. Apart from differences in natural radioactivity, there were no other significant differences between the underground test and the laboratory test. This suggests that natural radioactivity may have a substantial impact on the DCR of PMTs. 

It is necessary to further investigate the DCR of PMTs to determine the specific influence of radiation on the DCR. Since all tested PMTs exhibit an increase in DCR, it can be inferred that the cause of this increase originates from a common source.

\subsection{Shielding Experiment}\label{section2.2}
To further verify the impact of natural radioactivity on the DCR of PMTs and to compare the results with those obtained from the underground experiment, a shielding experiment was conducted in the LPMT testing surface laboratory. The primary objective of this experiment was to isolate and attenuate external radioactive background through radiation shielding measures, thereby observing changes in the PMTs' DCR under shielded conditions. This experiment was designed to enhance the credibility of our conclusions regarding the influence of natural radioactivity and to provide a comparative reference to the underground experimental results.

\begin{figure}[htbp]
    \centering
    \includegraphics[width=8.5cm,height = 5.00cm]{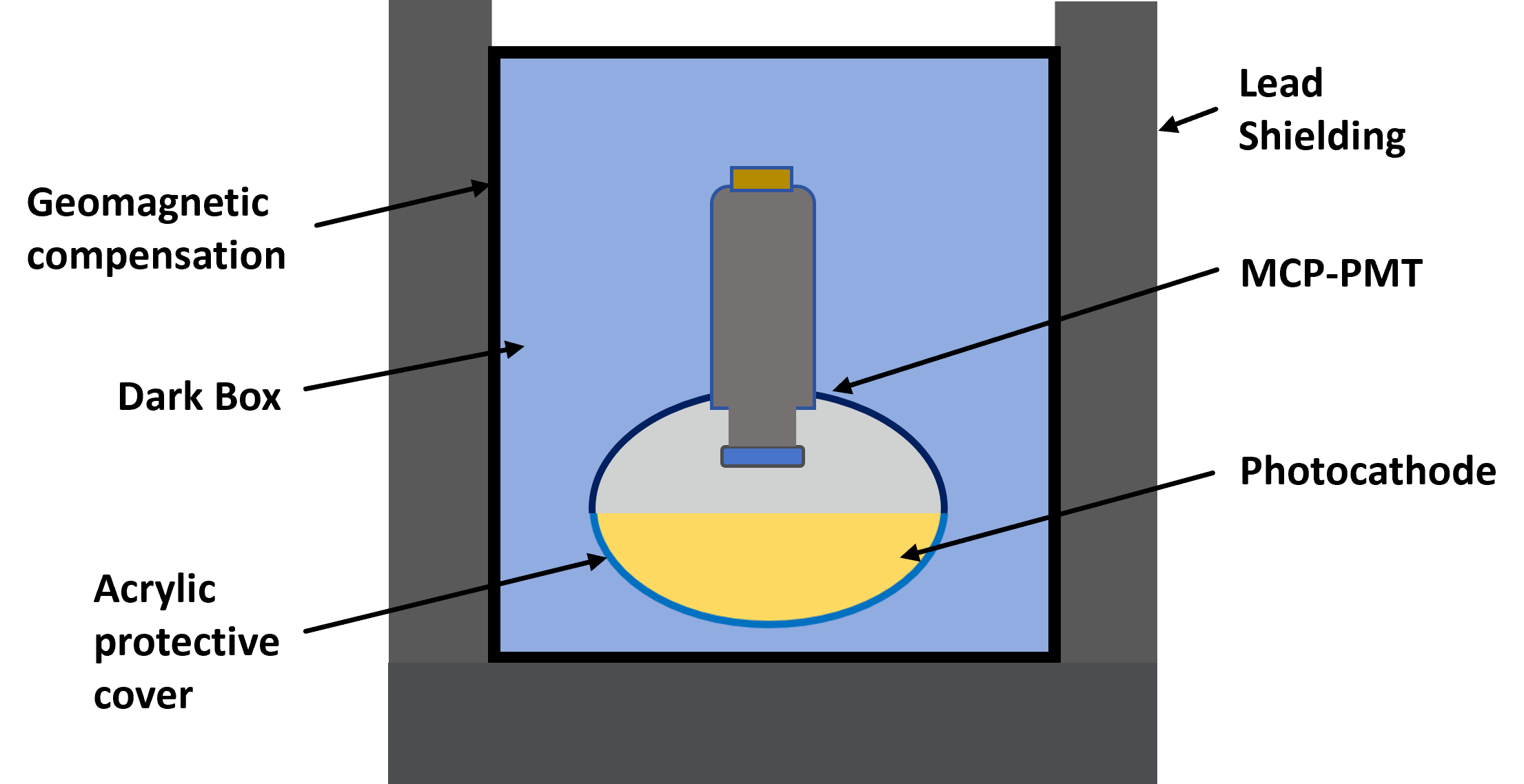}
    \qquad
    \includegraphics[width=4.67cm,height = 5.00cm]{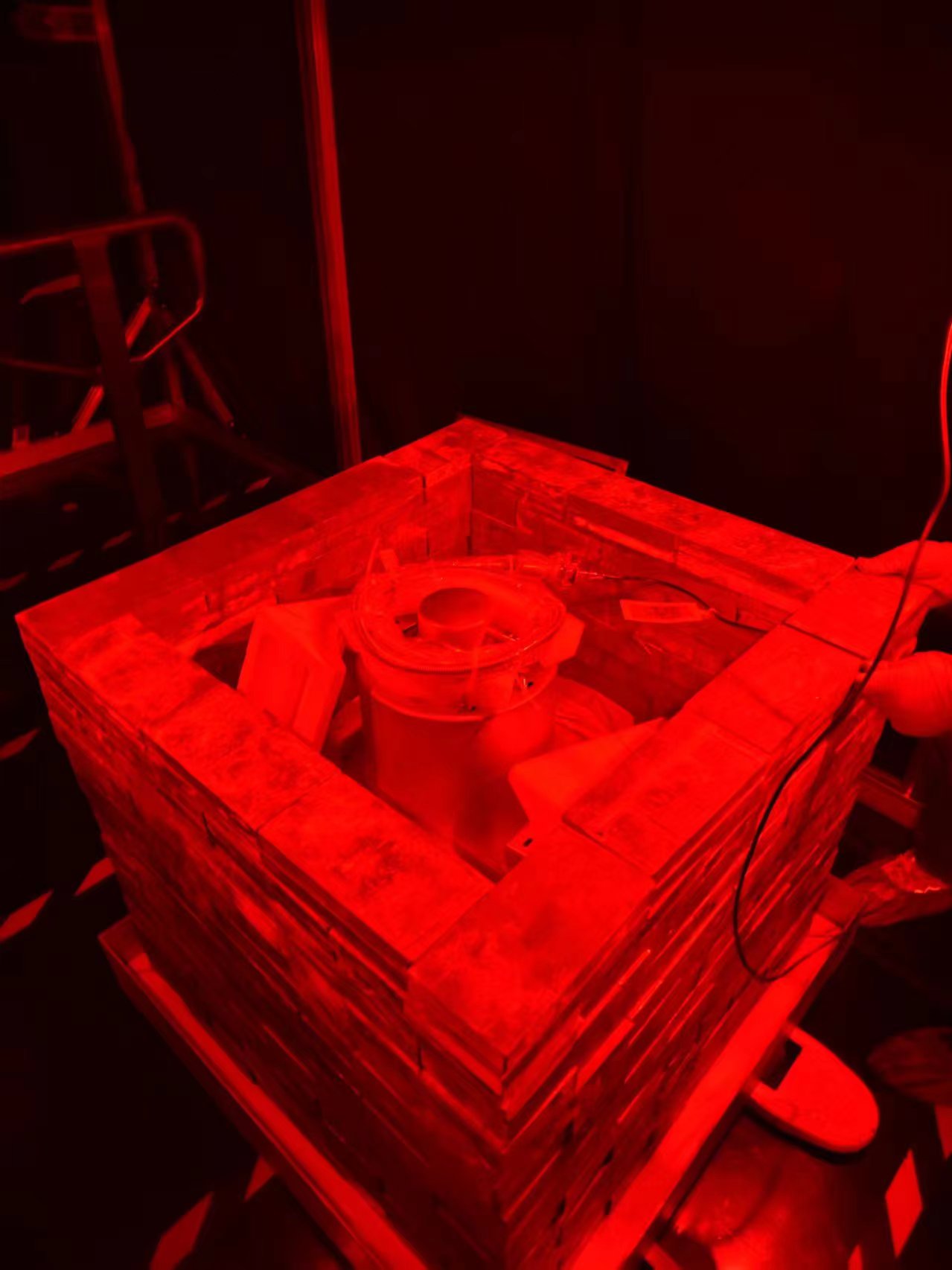}
    \caption{Shielding experiment setting up and the apparatus}
    \label{leading experiment}
\end{figure}

For testing a single PMT, we set up an experimental environment as shown in figure~\ref{leading experiment}. Specifically, we constructed a shielding structure using lead bricks to fit a 20-inch PMT. The dimensions of the lead bricks are $10~\mathrm{cm} \times 20~\mathrm{cm} \times 5~\mathrm{cm}$ and $10~\mathrm{cm} \times 20~\mathrm{cm} \times 2.5~\mathrm{cm}$, respectively. Two $2.5~\mathrm{cm}$ bricks were stacked to match the thickness of a single $5~\mathrm{cm}$ brick due to limited availability of the latter. After placing the PMTs inside this structure, we covered it with a black box containing geomagnetic compensation materials. When a PMT completes the cooling down process, this value is recorded as the intrinsic noise level under shielding conditions. All measurements were performed at an ambient temperature of $21^{\circ}\,\mathrm{C}$, identical to the nominal operating temperature of JUNO.

\begin{figure}[htbp]
    \centering
    \includegraphics[width=0.7\linewidth]{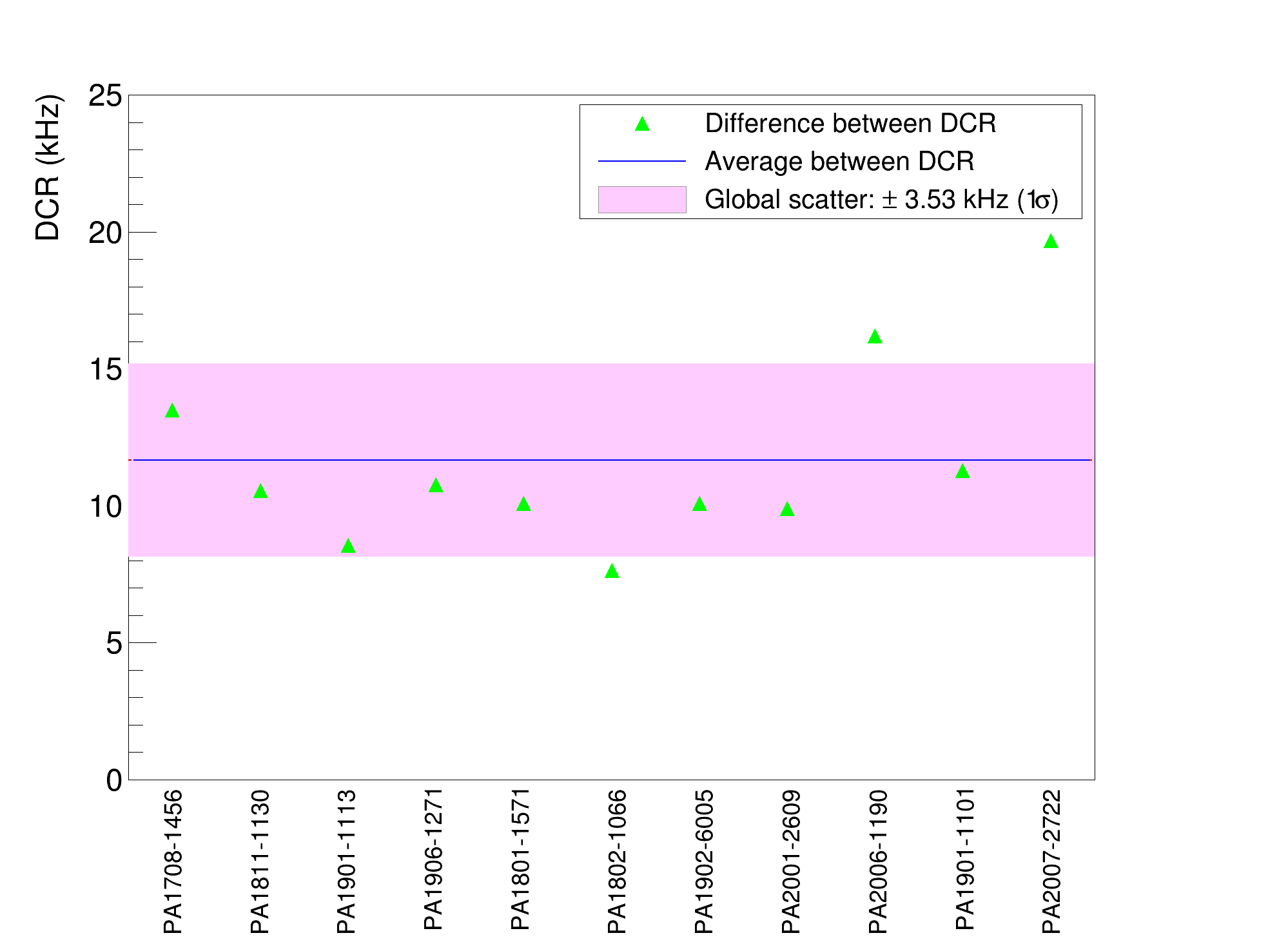}
    \caption{Shielding experiment results of MCP-PMTs under different configurations.}
    \label{Fig5}
\end{figure}

Subsequently, we took out the PMT from the lead shielding structure and retested it, keeping the PMT covered by the geomagnetic compensation black box in the same magnetic field intensity. Under these conditions, the PMT is influenced by external radioactive background, and its stabilized DCR reflects the noise level without lead shielding.
Under these conditions, we tested a total of 11 MCP-PMTs, with the distribution of test results displayed in table~\ref{tab:PMT_Shielding}.
\begin{table}[htbp]
  \centering
  \caption{PMTs shielding experiment results}
  \begin{tabular}{lccc}
    \toprule
    PMT Serial Number & \makecell{Before \\ shielding DCR (kHz)} & \makecell{After \\ shielding DCR (kHz)} & Difference(kHz)  \\
    \midrule
    PA1708-1456 & 46.19 & 32.70 & 13.49 \\
    PA1811-1130 & 74.46 & 63.90 & 10.56 \\
    PA1901-1113 & 17.77 & 9.20  & 8.57 \\
    PA1906-1271 & 25.23 & 14.47 & 10.76 \\
    PA1801-1571 & 29.79 & 19.70 & 10.09 \\
    PA1802-1066 & 32.53 & 24.90 & 7.63 \\
    PA1902-6005 & 59.20 & 49.11 & 10.09 \\
    PA2001-2609 & 29.94 & 20.05 & 9.89 \\
    PA2006-1190 & 45.67 & 29.46 & 16.21 \\
    PA1901-1101 & 26.11 & 14.82 & 11.29 \\
    PA2007-2722 & 48.74 & 29.05 & 19.69 \\
    The average & 39.60 & 27.94 & 11.66 \\
    \bottomrule
  \end{tabular}
  \label{tab:PMT_Shielding}
\end{table}
As shown, there is a notable difference in the DCR before and after shielding. For all tested PMTs, the average DCR difference is around 11 kHz as indicated by figure~\ref{Fig5},  with the global scatter representing the sample standard deviation.

These findings confirm that PMTs are influenced by radioactive factors. Considering that each step in the natural radioactive decay chains emits different types of radiation, including $\alpha$ particles, $\beta$ particles, and $\gamma$ rays, the shielding experiment results suggest the following conclusion: since $\alpha$ particles have a range of only a few centimeters in air, we infer that the DCR of PMTs is mainly affected by $\gamma$ rays or $\beta$ particles from natural radioactivity. This is because both $\gamma$ rays and $\beta$ particles have relatively strong penetrating abilities and longer ranges in air, allowing them to reach the PMT surface and contribute to the increase in DCR. 

The larger DCR variation of the PMTs in Table~\ref{tab:PMT_underground_exp} relative to Table~\ref{tab:PMT_Shielding} is mainly due to the use of different production batches in the two experiments. JUNO installed roughly 2000 20-inch PMTs, and inherent performance dispersion exists among individual PMTs. In the underground experiment, we deliberately used PMTs from JUNO pre-installation studies, which were never installed in the detector and were dedicated to investigating the impact of natural radioactivity on the DCR, in order to safeguard the PMT inventory and to minimize operational risks associated with extended handling and transport in the underground environment. In the shielding experiment, PMTs were deliberately selected to cover different DCR ranges in order to ensure the representativeness and robustness of the results. This intentional inclusion of devices with a broad spread in DCR leads to a larger error  in the average value. Moreover, in both experiments we observed a clear change in DCR even though different PMTs were used, so the use of different batches does not compromise our conclusion regarding the impact of environmental radioactivity on the PMTs' DCR. 

\subsection{ Radioactivity source experiment }

\subsubsection{\texorpdfstring{\(^{60}\mathrm{Co}\) irradiation experiment}{60Co irradiation experiment}}

In previous studies on the noise characteristics of MCP-PMTs, we found that natural radioactivity has a significant impact on the DCR. To further investigate the specific effects of different types of radiation on DCR, we built a small PMT testing system at the Institute of High Energy Physics in Beijing to measure changes in the DCR and charge spectrum of PMTs under the influence of various types of radiation.

In the experiment, the PMT was placed inside a dark box covered externally with geomagnetic compensation material, with an additional layer of black cloth to block ambient light, the same setup as we did at the JUNO site. First, we examined the impact of gamma radiation on the DCR of the PMT, using a \(^{60}\mathrm{Co}\) source to generate $\gamma$ rays\cite{genna1955absolute}.

\begin{figure}
    \centering
    \includegraphics[width=0.7\linewidth]{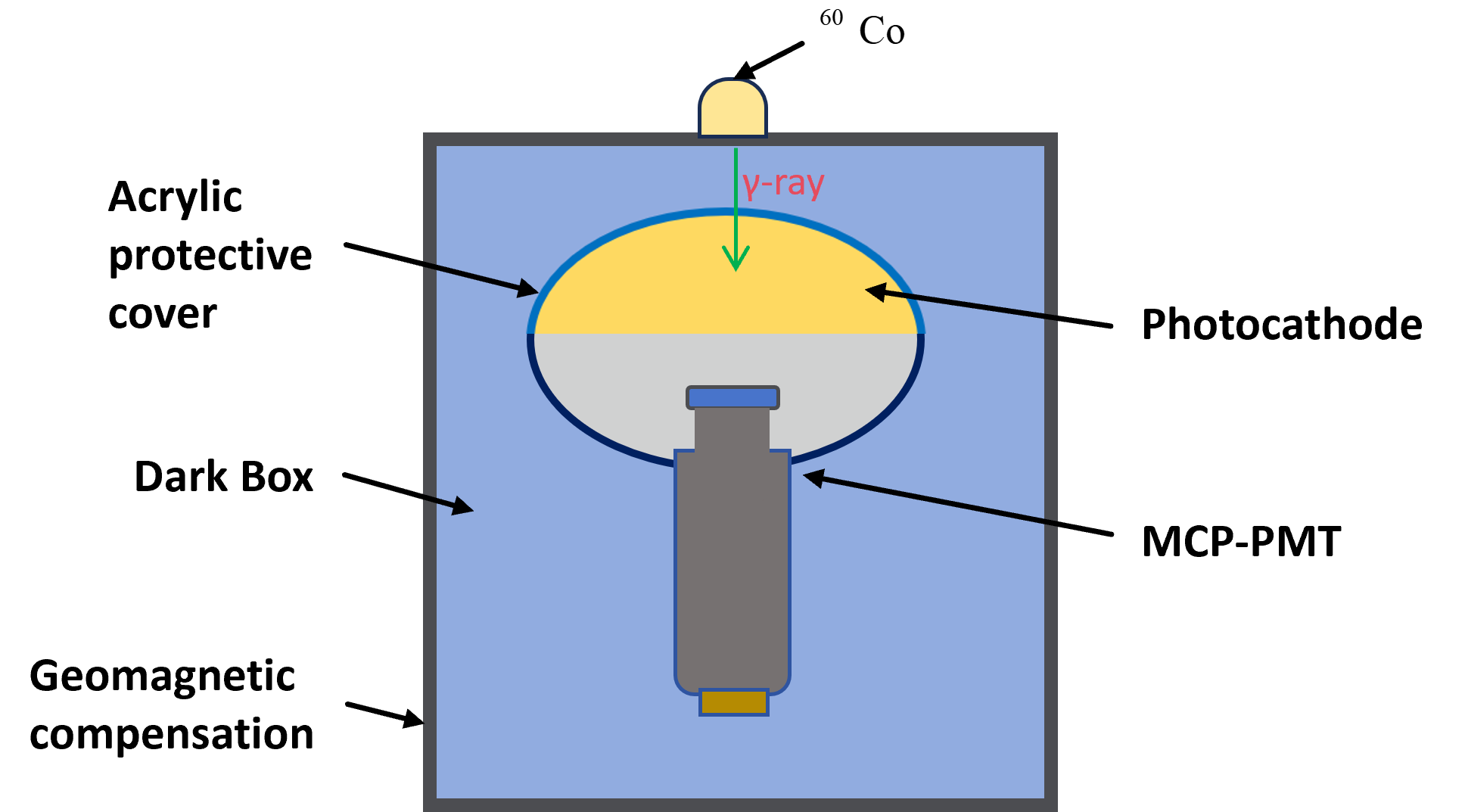}
    \caption{\(^{60}\mathrm{Co}\) experiment Setting up}
    \label{Co-60setup}
\end{figure}

\begin{figure}[htbp]
    \centering
    \includegraphics[width=0.45\linewidth]{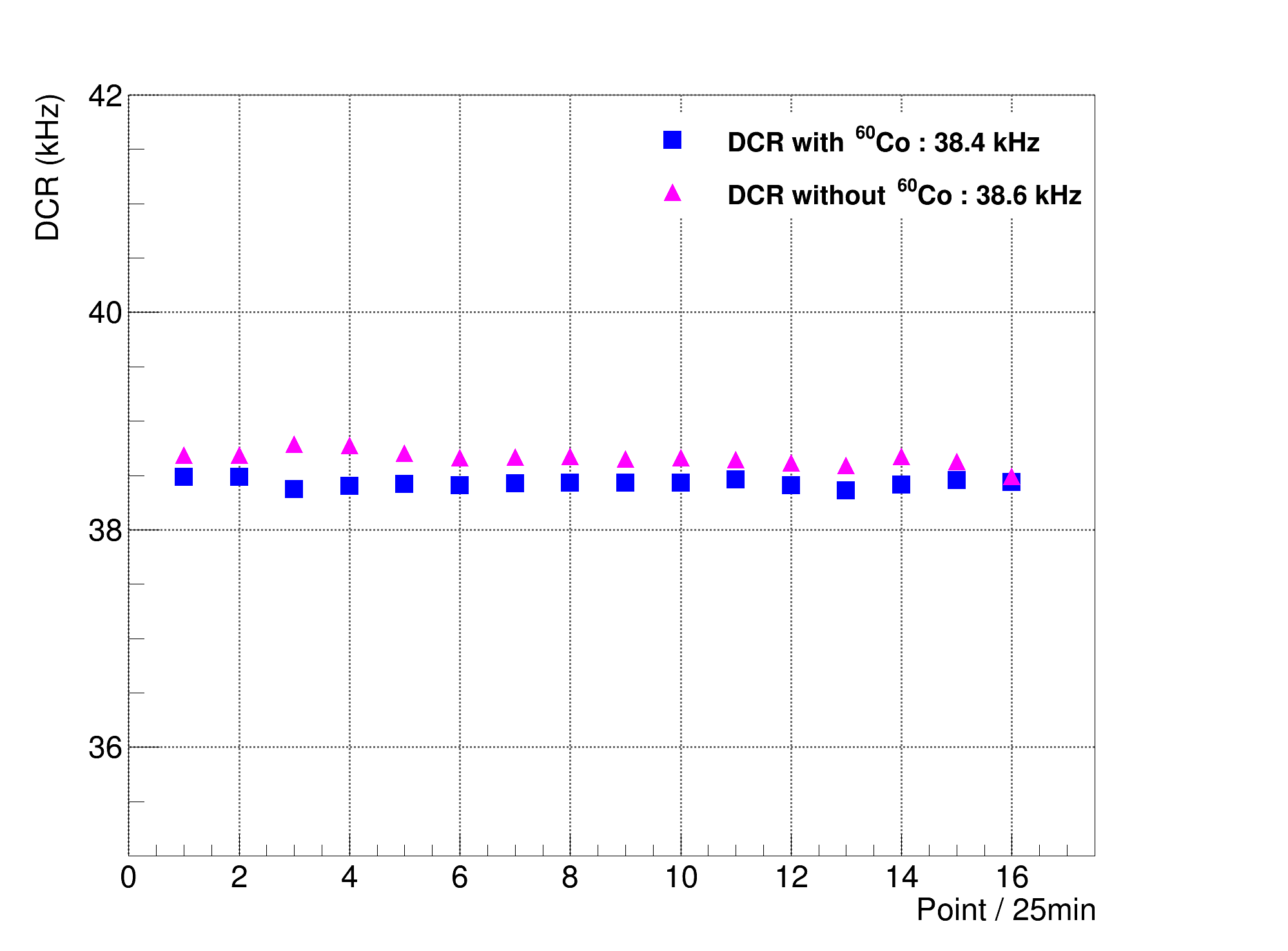}
    \qquad
    \includegraphics[width=0.45\linewidth]{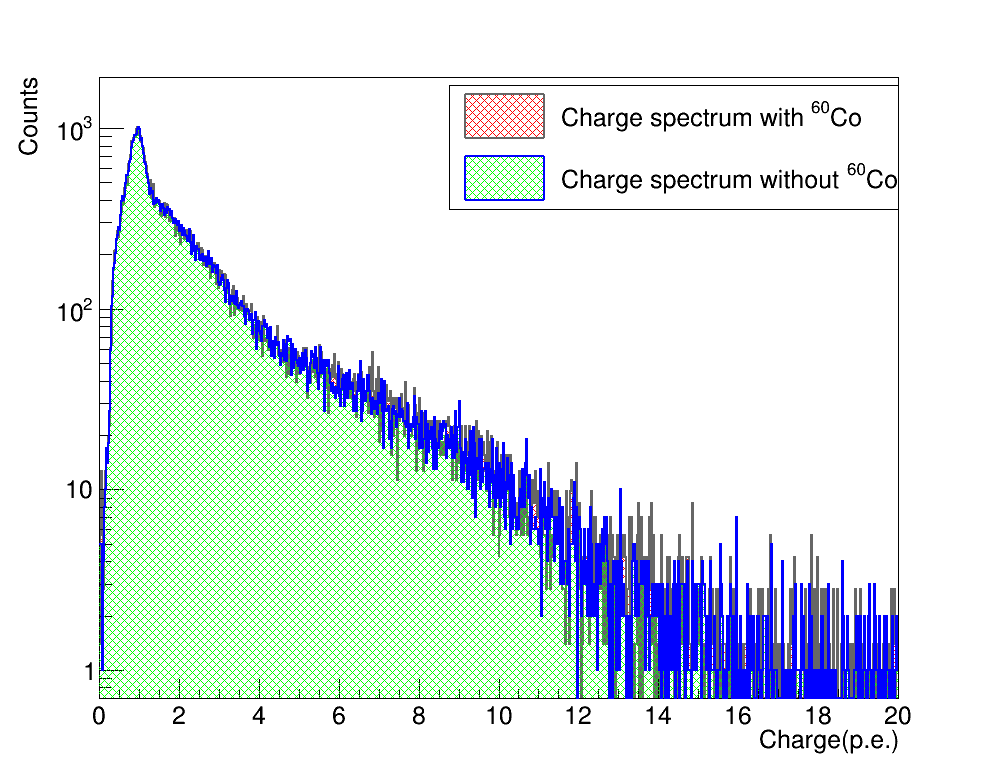}
    \label{fig:Co—60 charge spectrum}
    \caption{The DCR of \(^{60}\mathrm{Co}\) experimental result (left) and charge spectrum of \(^{60}\mathrm{Co}\) irradiation experiment (right)}
    \label{fig:Co60_combined}
\end{figure}

The \(^{60}\mathrm{Co}\) source was placed in a lead box, where $\gamma$ rays generated from its decay were collimated through an aperture in the lead box. As shown in figure~\ref{Co-60setup}, the lead box was positioned on the black cloth, and the collimator hole was aligned with the photocathode glass section of the PMT inside the dark box, allowing the emitted $\gamma$ rays to fully enter the vacuum region of the PMT. This setup enabled precise measurement of the $\gamma$ radiation’s effect on the DCR.
In this experiment, we tested a PMT, and the measured DCR spectrum and charge spectrum are shown in figure~\ref{fig:Co60_combined}.

Test results indicate that, in the absence of \(^{60}\mathrm{Co}\) irradiation, the PMT exhibits a DCR of 38.6 kHz, with its charge spectrum showing a peak at the single-electron level followed by exponential decay—characteristic of this PMT model’s typical charge distribution. Following \(^{60}\mathrm{Co}\) irradiation and subsequent cooling, the DCR slightly decrease to 38.4 kHz, while the charge spectrum still displays a peak at the single-electron level, with no noticeable differences from the non-irradiated spectrum. This change in DCR falls within the range of expected random fluctuations, suggesting that \(^{60}\mathrm{Co}\) irradiation has minimal impact on the DCR of the PMTs.

\subsubsection{\texorpdfstring{\(^{55}\mathrm{Fe}\)}{55Fe} irradiation experiment}

\begin{figure}
    \centering
    \includegraphics[width=0.7\linewidth]{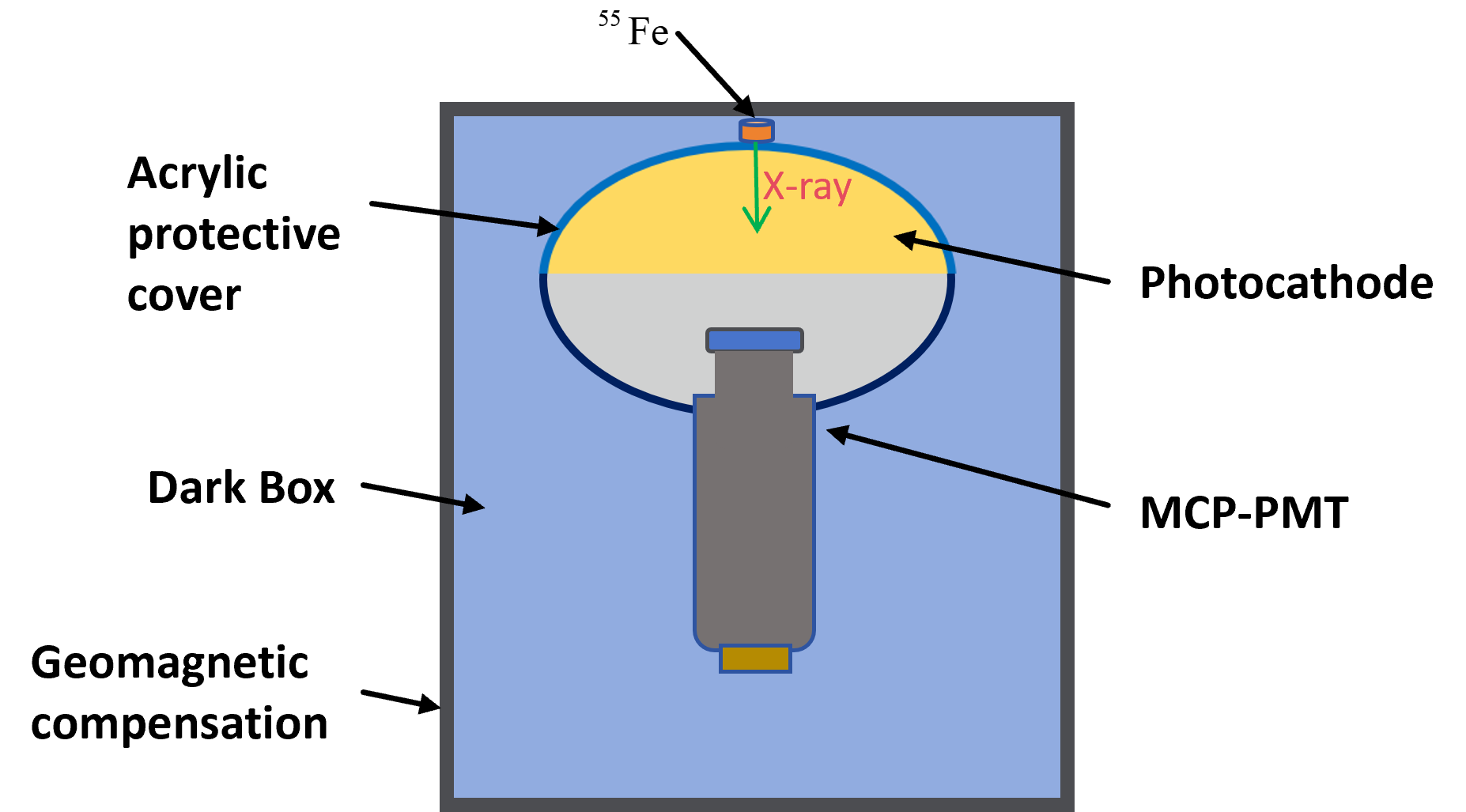}
    \caption{$^{55}\mathrm{Fe}$ experiment Setting up}
    \label{Fe-55setup}
\end{figure}

Considering that \(^{60}\mathrm{Co}\) emits high-energy gamma rays with energies of $1.17~\mathrm{MeV}$ and $1.33~\mathrm{MeV}$, the radioactive source was replaced with \(^{55}\mathrm{Fe}\) to investigate the effect of low-energy photons on the DCR. \(^{55}\mathrm{Fe}\) primarily emits low-energy $\mathrm{X}$-rays at $5.9~\mathrm{keV}$ and $6.49~\mathrm{keV}$. In the experiment, the \(^{55}\mathrm{Fe}\) source was affixed to the top of the PMT using transparent adhesive tape to ensure that the emitted X-rays could effectively enter the interior of the PMT, while all other experimental conditions were kept unchanged. The experimental setup is shown in figure ~\ref{Fe-55setup}.

 Test results indicate that, in the absence of \(^{55}\mathrm{Fe}\) radiation, the PMT exhibited a DCR of 38.6 kHz and its charge spectrum displayed the characteristic single-electron peak. Under \(^{55}\mathrm{Fe}\) irradiation, the PMT’s DCR increased to 41.0 kHz, yet the overall shape of the charge spectrum remained consistent with the non-irradiated condition, with no new peak observed in the high-charge region. 
\begin{figure}[htbp]
    \centering
    \includegraphics[width=0.45\linewidth]{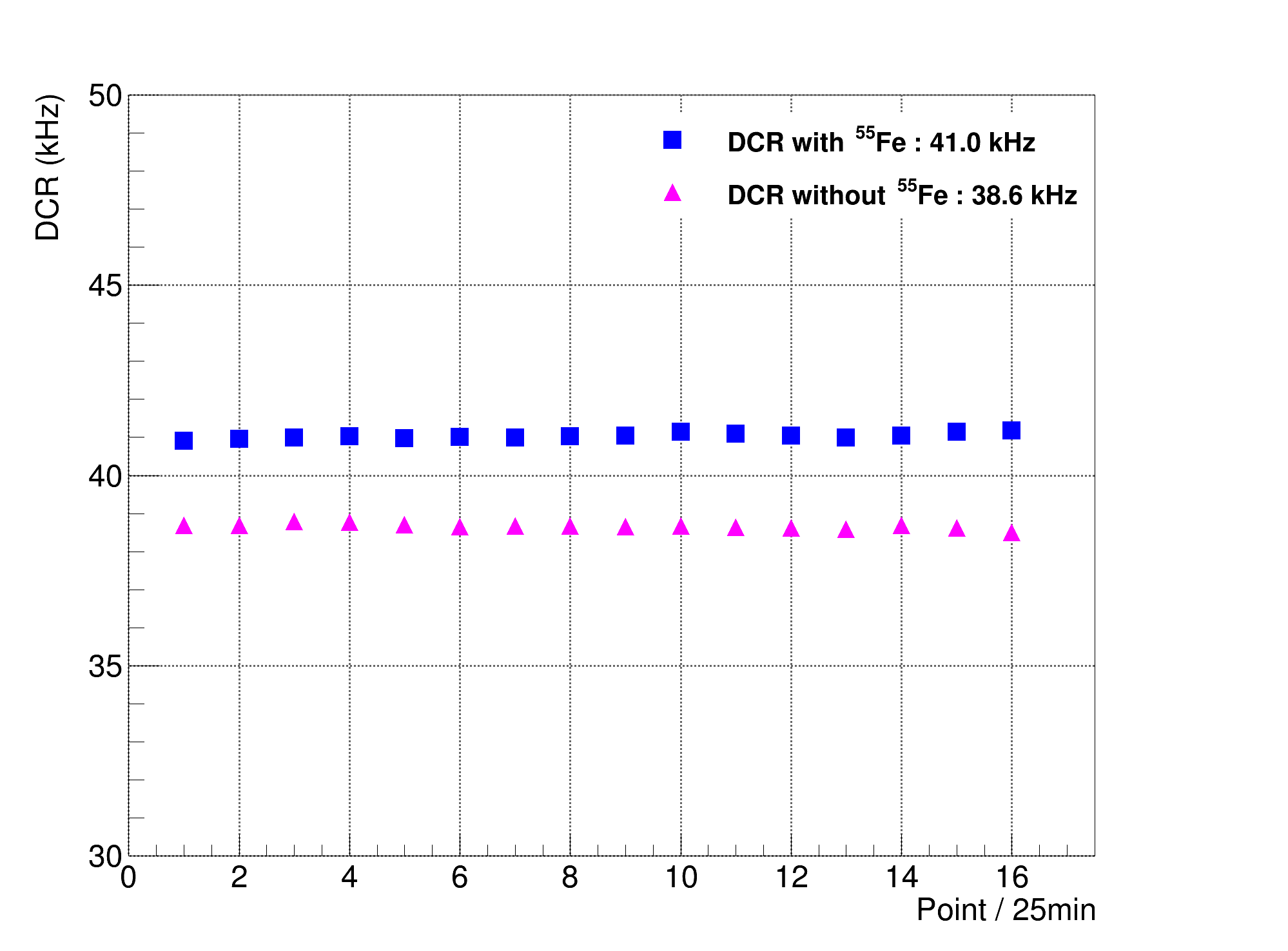}
    \qquad
    \includegraphics[width=0.45\linewidth]{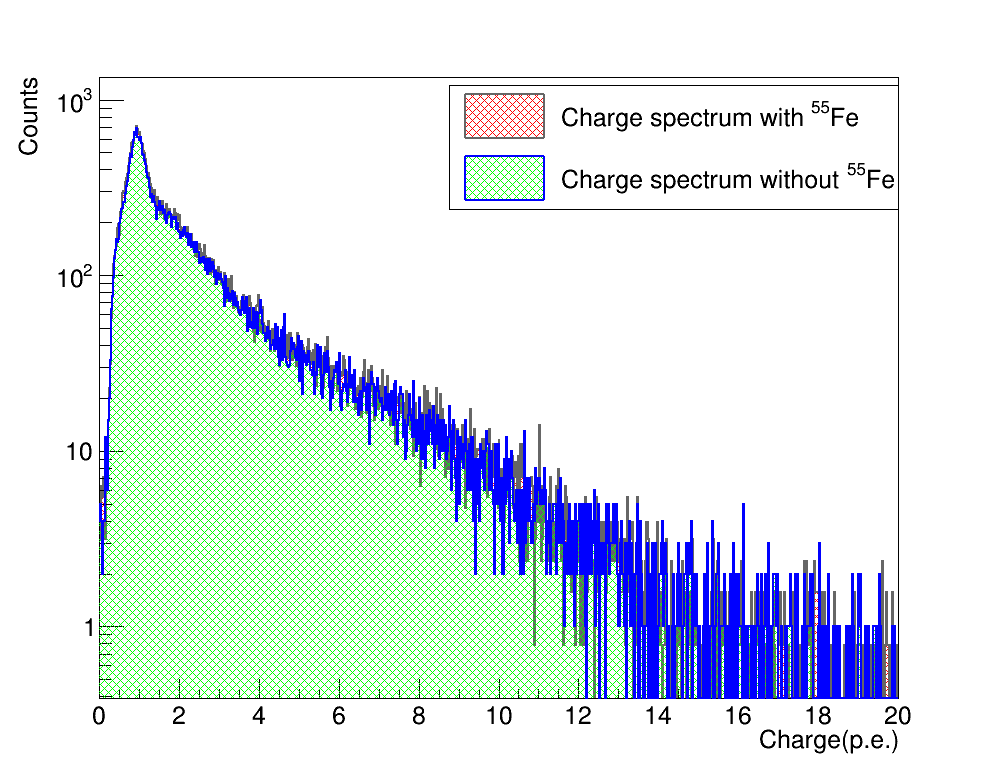}
    \caption{The DCR of \(^{55}\mathrm{Fe}\) experimental result (left) and charge spectrum of \(^{55}\mathrm{Fe}\) irradiation experiment (right)}
    \label{fig:Fe55_combined}
\end{figure}
In the measured DCR results, the average value under irradiation was approximately $2.4~\mathrm{kHz}$ higher than that in the non-irradiated condition. The charge spectra obtained under both conditions were nearly identical, exhibiting no significant differences. In contrast, The experiments conducted in sections~\ref{section2.1} and \ref{section2.2} showed average DCR variations exceeding $10~\mathrm{kHz}$, indicating that the overall impact of \(^{55}\mathrm{Fe}\) on DCR is relatively minor.

\subsubsection{\texorpdfstring{$^{90}\mathrm{Sr}$}{90Sr} irradiation experiment}

\begin{figure}[htbp]
    \centering
    \includegraphics[width=0.64\linewidth]{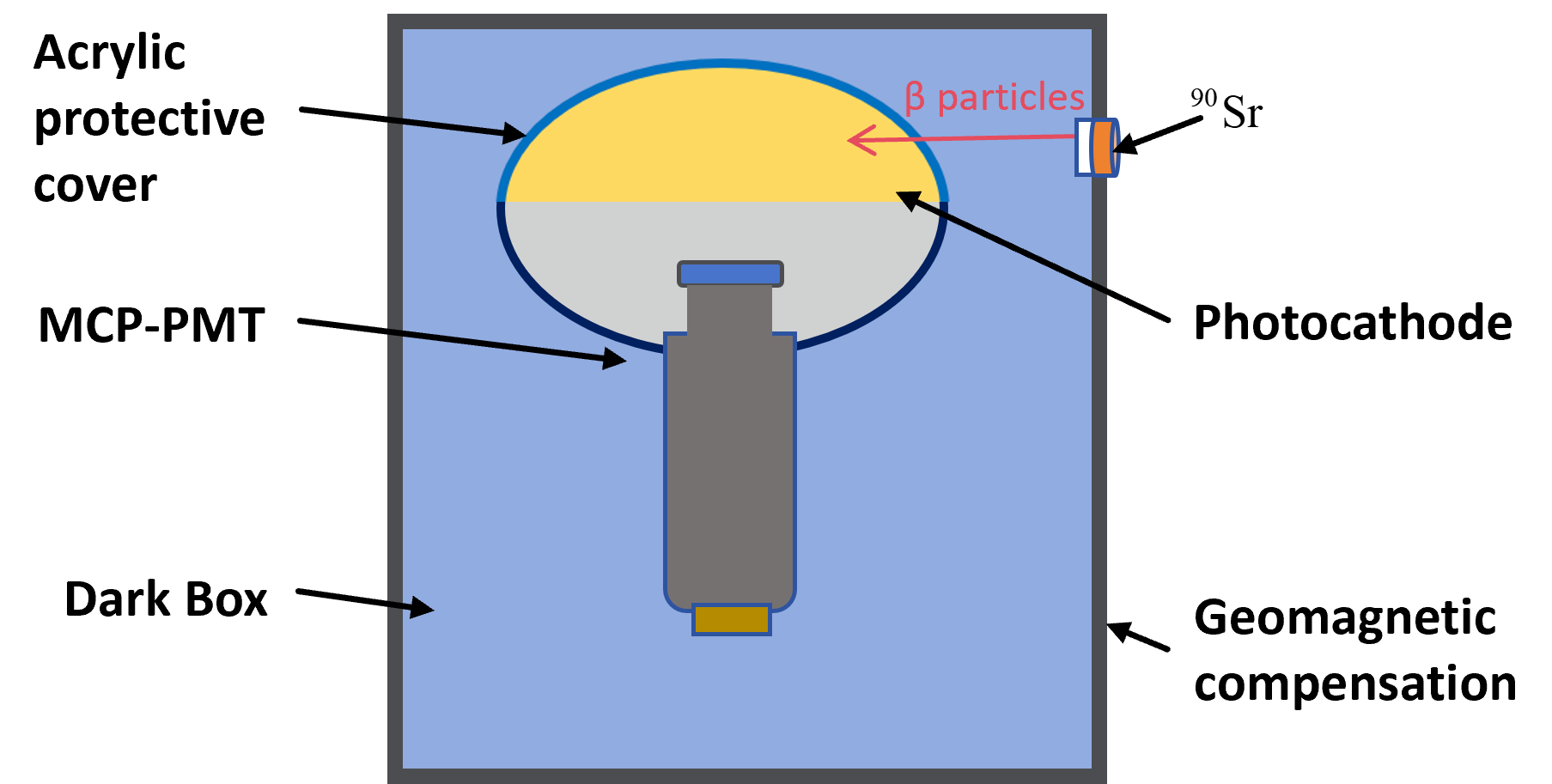}
    \qquad
    \includegraphics[width=0.27\linewidth]{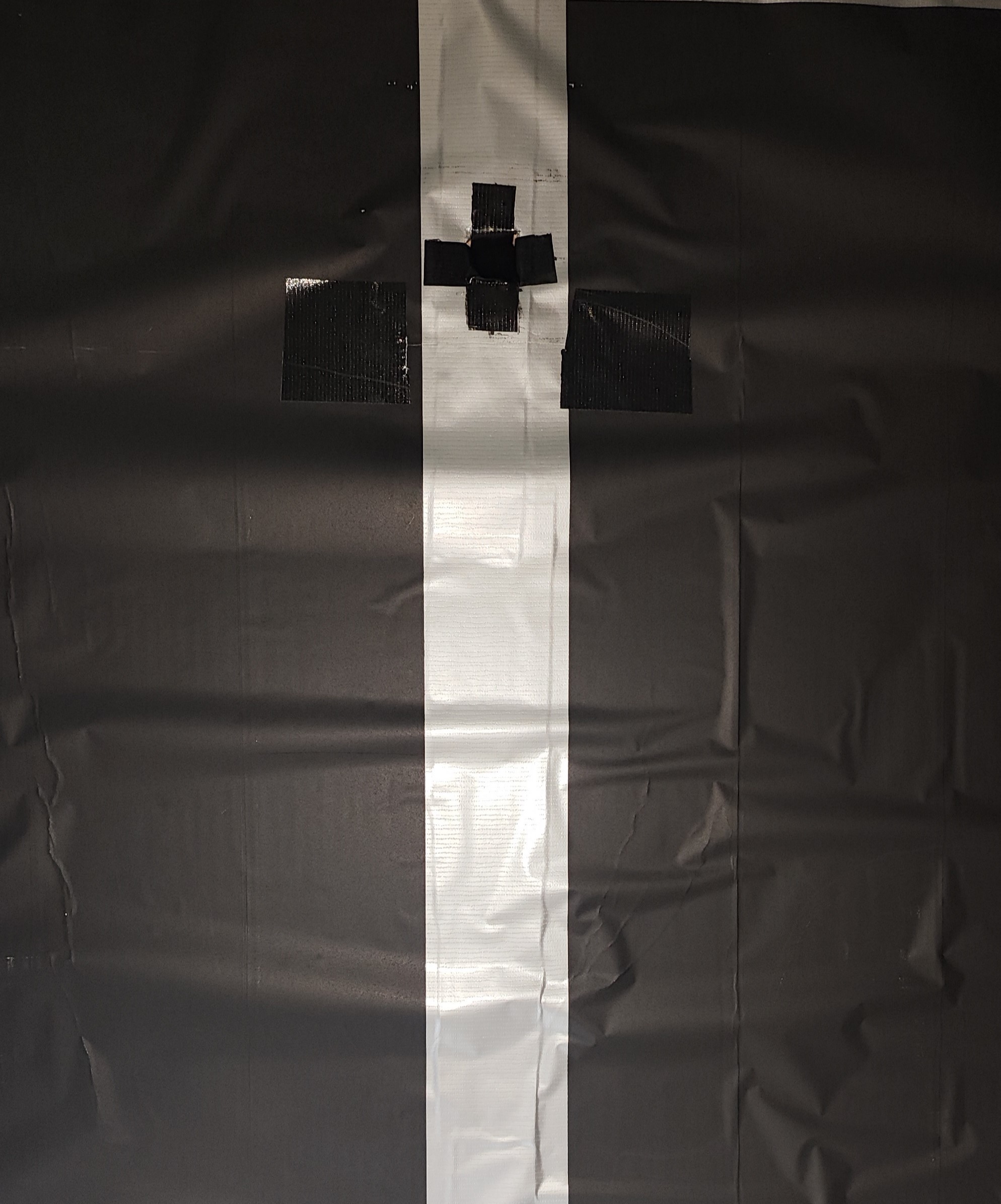}
    \caption{\(^{90}\mathrm{Sr}\) experiment Setting up (left) and Transmission window on the geomagnetic compensation material (right).}
    \label{fig:Sr90_combined}
\end{figure}

In the previous irradiation experiments, we found that $\gamma$ rays and $\mathrm{X}$-rays were not the primary contributors to the increase in DCR. However, natural radioactive decay not only emits $\gamma$ rays but also produces $\beta$ particles, which may play a more critical role in the observed changes in DCR. To investigate the effect of $\beta$ particles on the DCR, a \(^{90}\mathrm{Sr}\) radioactive source was selected as radiation source. In the experiment, the container holding the \(^{90}\mathrm{Sr}\) source was secured to the outer surface of the geomagnetic compensation box using adhesive tape. The experimental setup is shown in figure ~\ref{fig:Sr90_combined}.

\begin{figure}[htbp]
    \centering
    \includegraphics[width=7.15cm,height = 6.3cm]{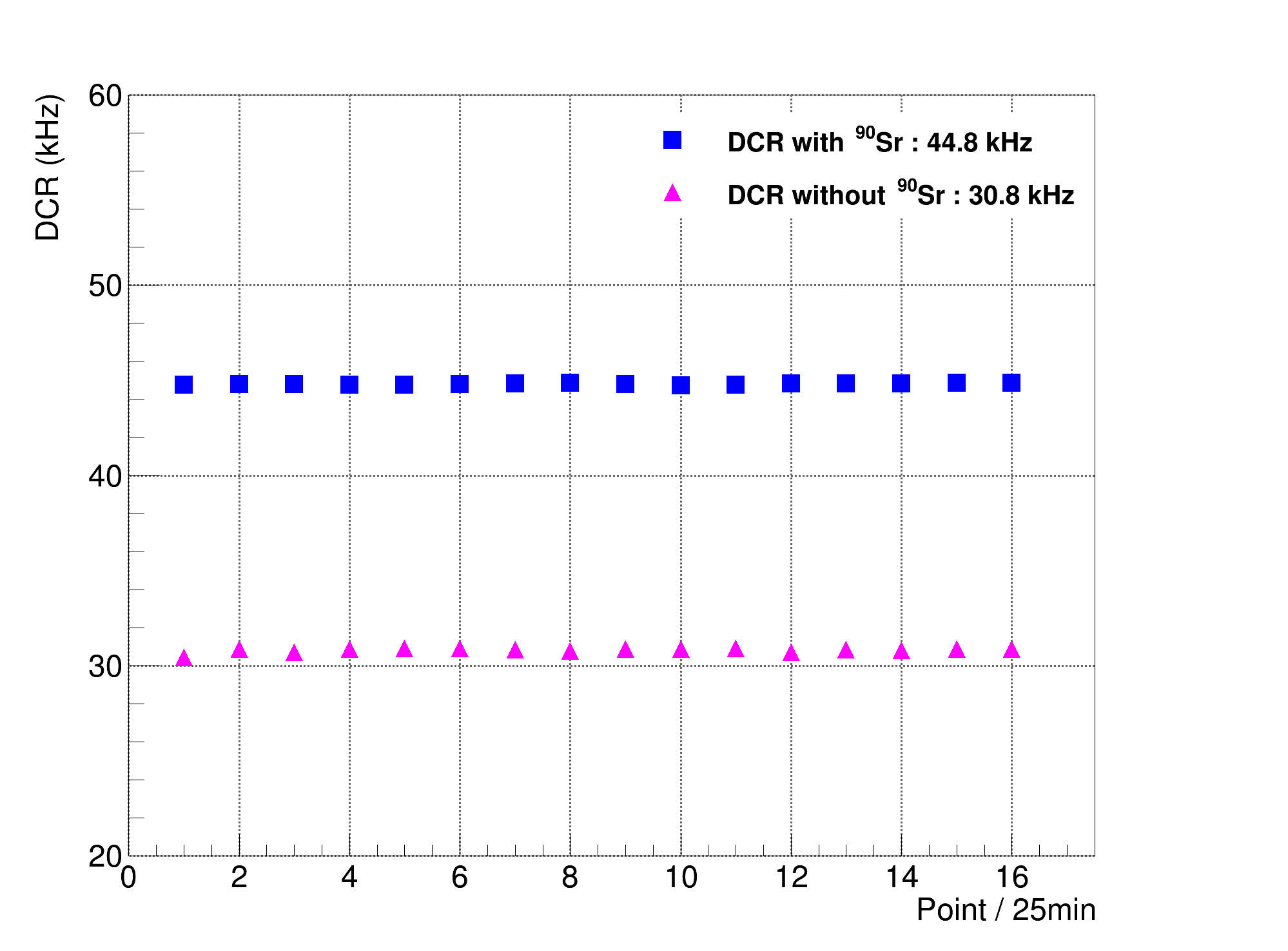}
    \qquad
    \includegraphics[width=7.1cm,height = 6.00cm]{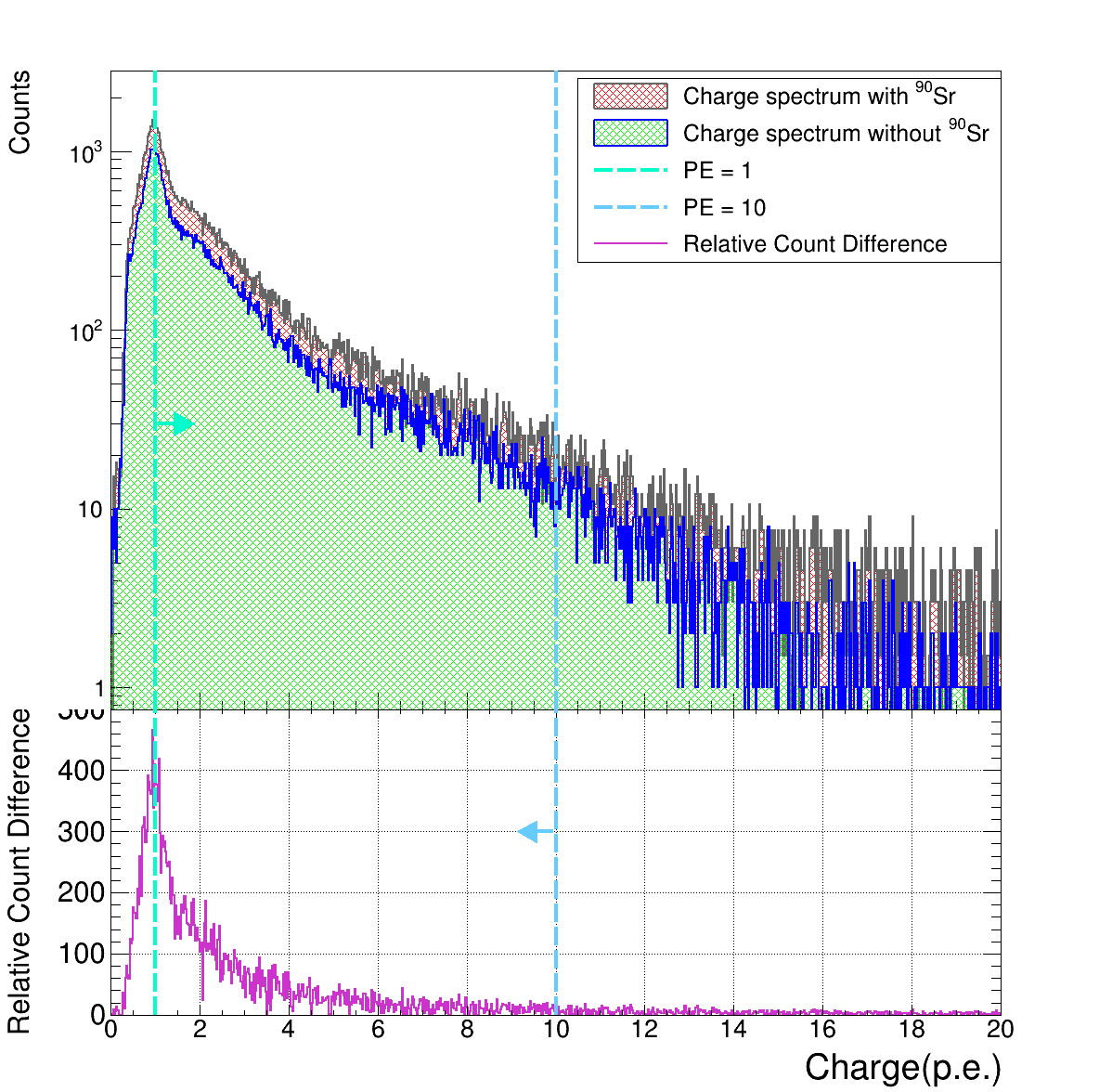}
    \caption{The DCR of \(^{90}\mathrm{Sr}\) experimental result (left) and charge spectrum of \(^{90}\mathrm{Sr}\) irradiation experiment (right)}
    \label{fig:Sr90_test_result}
\end{figure}

Considering that $\beta$ particles have weaker penetration capability compared to $\gamma$-rays, as shown in figure~\ref{fig:Sr90_combined}, a window was opened in the corresponding region of the geomagnetic compensation box to minimize the absorption and scattering of $\beta$ particles by the material. As shown in figure~\ref{fig:Sr90_test_result}, the DCR test results clearly indicate that when the PMT is exposed to the \(^{90}\mathrm{Sr}\) radioactive source, DCR increases significantly, with an average rise of approximately $14~\mathrm{kHz}$. According to the charge spectrum results, the curves under the two conditions do not overlap. Under \(^{90}\mathrm{Sr}\) irradiation, the number of detected electrons is noticeably higher than in the non-irradiated condition, with this increase primarily concentrated around the single photoelectron peak. Although in the $1 < \mathrm{PE} < 10$ range the electron count under irradiation remains higher than that without irradiation, the difference gradually diminishes beyond the single photoelectron region. Based on the combined analysis of the DCR test results and the charge spectrum distributions, it is evident that the $\beta$ particles emitted by \(^{90}\mathrm{Sr}\) significantly increase the DCR of the PMTs. The output signals primarily manifest as single-photoelectron events, similar to those induced by thermionic emission. 

In the JUNO experiment, every 20-inch PMT is fitted with a transparent and hemispherical acrylic protective cover (APC) whose primary purpose is to prevent implosion under the high hydrostatic pressure and thereby ensure operational safety. For the APC, the cover exhibits a light transmittance of $98.1\%$, an average thickness of $10~\mathrm{mm}$, and a refractive index of $n = 1.49$\cite{He_2024,He_2023}. It fully encloses the PMT's photocathode so as to maximise the transmission of incident photons. The PMTs employ a hybrid photocathode geometry, with a transmission bi-alkali (Sb–K–Cs) photocathode on the upper hemisphere and a reflective photocathode on the lower hemisphere. The effective response wavelength range of the photocathode is approximately 300$~\mathrm{nm}$ to 690$~\mathrm{nm}$. The $\beta$ particles emitted by the \(^{90}\mathrm{Sr}\) source have a maximum energy of $2.3~\mathrm{MeV}$ and follow a continuous energy spectrum\cite{yalcin2011analytical}, indicating that they are capable of generating Cherenkov radiation in both the photocathode glass and the APC of the PMTs. A fraction of this Cherenkov emission lies in the several hundred nanometer range, which is within the photocathode’s effective sensitivity and these photons can reach the photocathode, produce photoelectrons via the photoelectric effect.

At the same time, in this experiment, the \(^{90}\mathrm{Sr}\) source was fixed at the side of the PMT, whereas the \(^{60}\mathrm{Co}\) and \(^{55}\mathrm{Fe}\) sources were placed on top. Although the lateral placement differs geometrically from the top configuration, its overall influence on the variation in DCR is negligible. Because Cherenkov radiation forms a light cone with its apex at the emission point, this orientation changes only the incidence positions of the photons on the photocathode and does not appreciably alter the total number of photons.

Considering that variations in DCR were also observed in both the underground and lead-shielding experiments, we further suggest that $\beta$ particles emitted from radionuclides in the surrounding rocks of the underground environment such as \(^{228}\mathrm{Ra}\) and \(^{234}\mathrm{Th}\) which undergo $\beta$ decay and are produced by radioactive decay chains are likely major contributors to the elevated DCR observed in the PMTs. Conversely, the lead shielding effectively suppresses the DCR by attenuating these $\beta$ particles, thereby reducing the DCR of PMTs inside the shielded box.

\section{Geant4 simulation}
In the previous sections, we confirmed that the DCR of PMTs is influenced by natural radioactive backgrounds by performing shielding tests and conducting measurements in an underground environment. Subsequently, we conducted irradiation experiments using a radioactive source and observed that $\beta$ particles are the primary contributors to the observed increase in DCR. 

At an average thickness of \(10~\mathrm{mm}\), the $\beta$ particles emitted by \(^{90}\mathrm{Sr}\) cannot penetrate the APC effectively. In addition, some of the Cherenkov photons generated lie within the response range of the PMT photocathode. Therefore, we hypothesize that the increase in DCR observed in the experiment is primarily attributable to the contribution of Cherenkov photons. 

\begin{figure}[htbp]
    \centering
    \subfigure[MCP-PMT dimensions]{
        \includegraphics[width=0.36\linewidth]{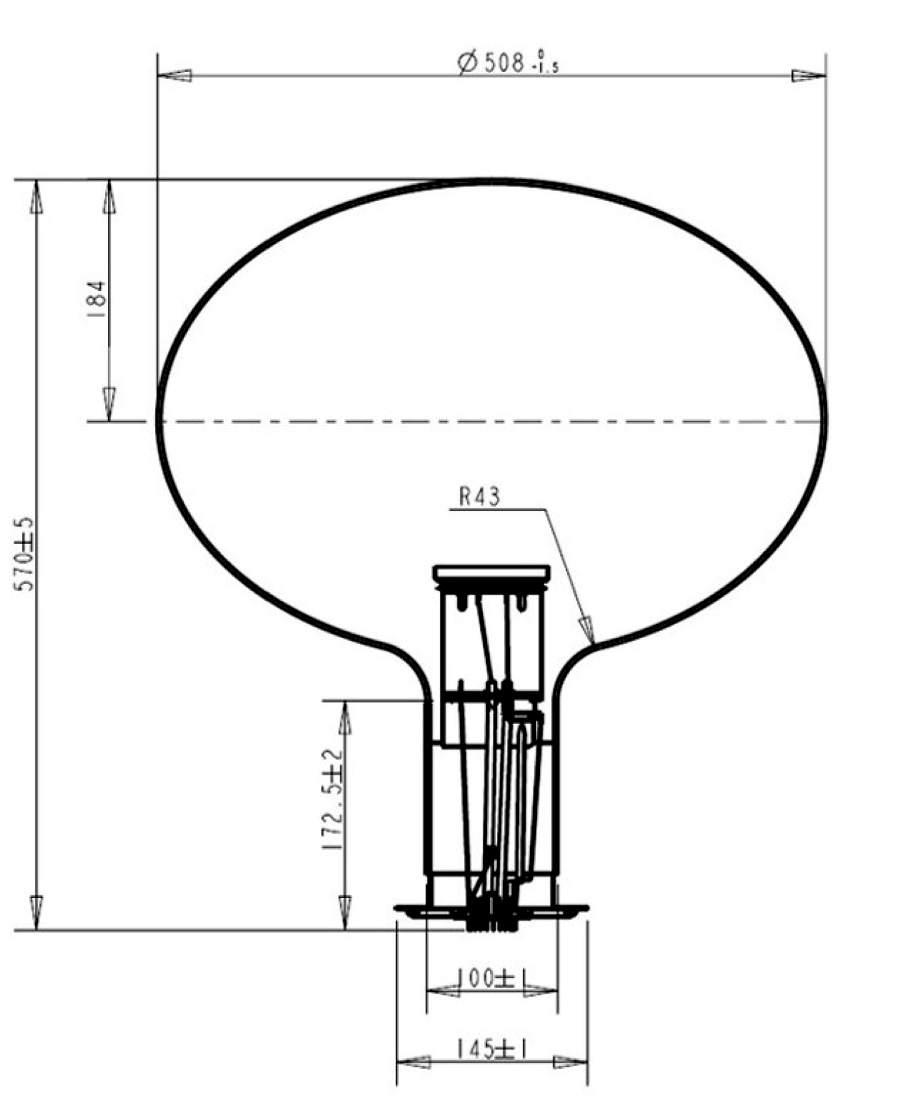}
        \label{fig:MCP-PMT-size-detail}
    }
    \qquad
    \subfigure[MCP-PMT after encapsulation]{
        \includegraphics[width=0.37\linewidth]{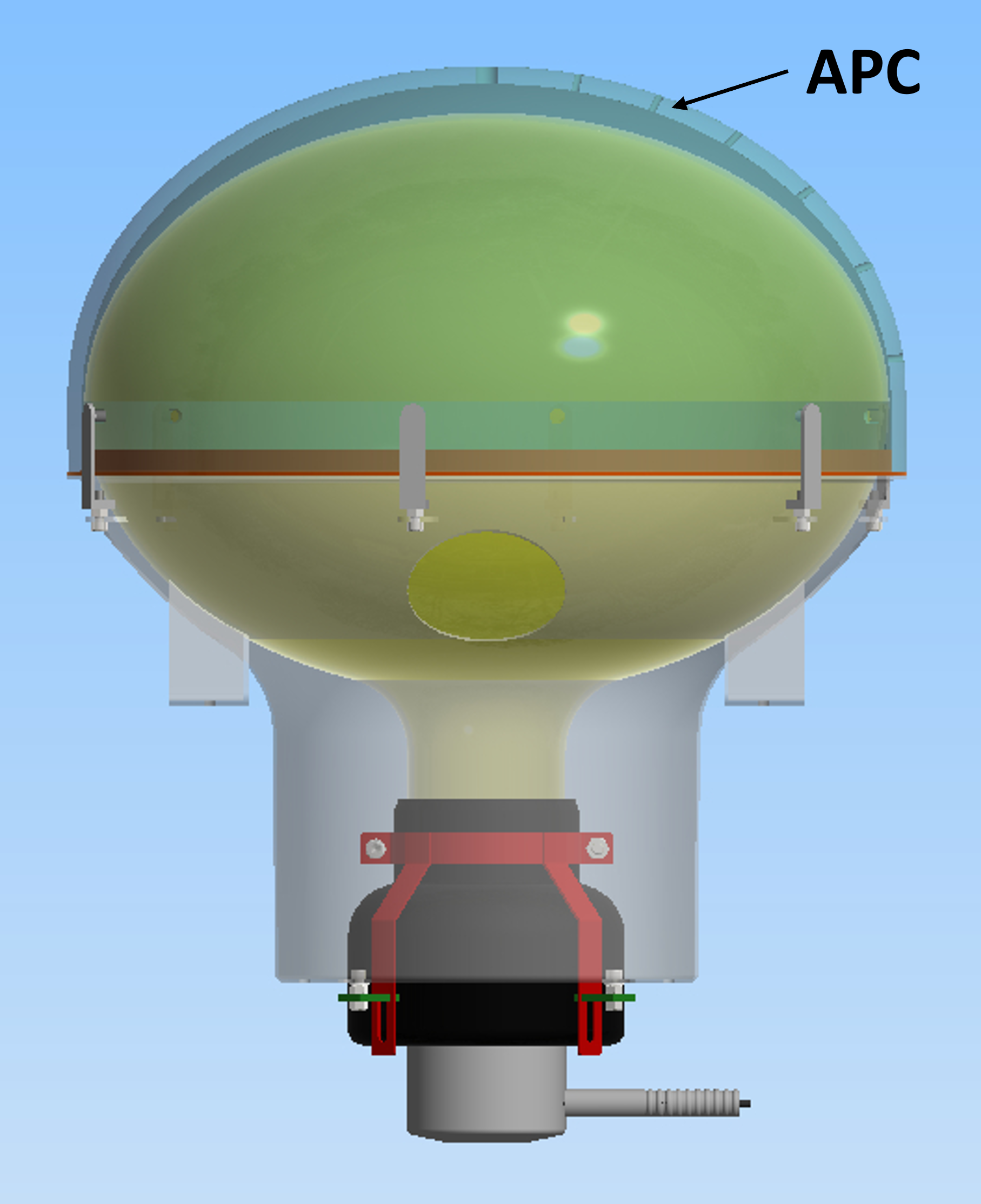}
        \label{fig:MCP-PMT-cover}
    }
    \caption{ The left image (a) depicts the detailed dimensions of the MCP-PMT. The right image (b) shows the appearance of the fully encapsulated MCP-PMT, with the light green region indicating the acrylic protective cover used.\cite{He_2024,He_2023}.}
    \label{fig:MCP-PMT-overview}
\end{figure}

To validate this hypothesis, we developed a detailed Geant4-based simulation model of a 20-inch MCP-PMT. This model was specifically designed to evaluate the effects of $\beta$ particles on the PMT performance. Figure~\ref{fig:MCP-PMT-overview} shows the specific dimensions of the PMT used in both the experiment and the Geant4 simulation.

\begin{figure}
    \centering
    \includegraphics[width=0.4\linewidth]{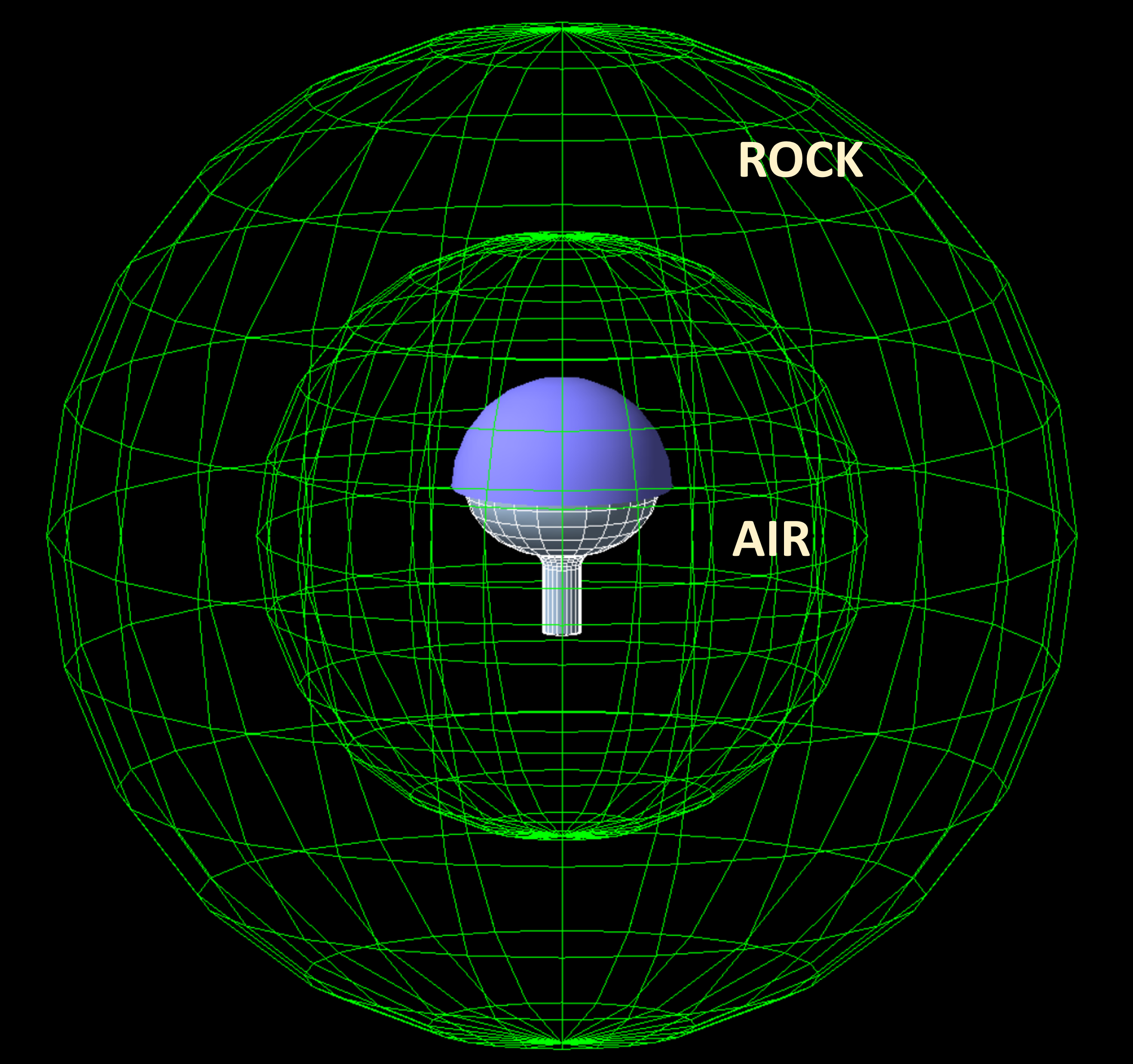}
    \caption{PMT simulation model}
    \label{fig:geant4-sim}
\end{figure}

Figure~\ref{fig:geant4-sim} illustrates the modeled structure of the MCP-PMT as implemented in Geant4.
 The simulation environment was configured to replicate realistic experimental conditions. The PMT was placed within a simulated atmospheric composition and enclosed by a hollow spherical shell. The PMT was precisely positioned at the center of the sphere. The purpose of introducing the spherical shell in the simulation is to facilitate particle trace reconstruction in the subsequent data analysis. The blue region in the figure represents the APC modeled in Geant4. Furthermore, in the simulation we omitted any radioactive elements in the air and surrounding rock to prevent the PMT from being affected by natural radioactivity.

 To ensure that the simulation accurately reflects the experimental conditions, the emission point of the $\beta$ particles was set at a position $125~\mathrm{mm}$ from the equator of the MCP-PMT model and $100~\mathrm{mm}$ above it, with the emission direction fixed along the negative $\mathrm{X}$-axis. At the same time, the $\beta$ particles  energy spectrum, we adopted the theoretical results reported in Ref.\cite{yalcin2011analytical}. The particle energies were sampled from the $^{90}\mathrm{Sr}-^{90}\mathrm{Y}$ decay spectrum model, and an interpolation procedure was applied to generate an energy distribution suitable for the Geant4 simulation.

In this Geant4 simulation, a total of one million $\beta$ particles were emitted, and each photon successfully entering the interior of the PMT was counted as a DCR signal. Considering that the MCP-PMTs used in JUNO are sensitive to photons within the wavelength range of $300~\mathrm{nm}$ to $690~\mathrm{nm}$, a wavelength selection was applied in the simulation to retain only photons within this range. Additionally, only the resulting photons weighted by the average detection efficiency of 28\%, corresponding to the global average over the MCP-PMTs surface used in the JUNO experiment, were considered as valid DCR signals.

\begin{figure}
    \centering
    \includegraphics[width=0.9\linewidth]{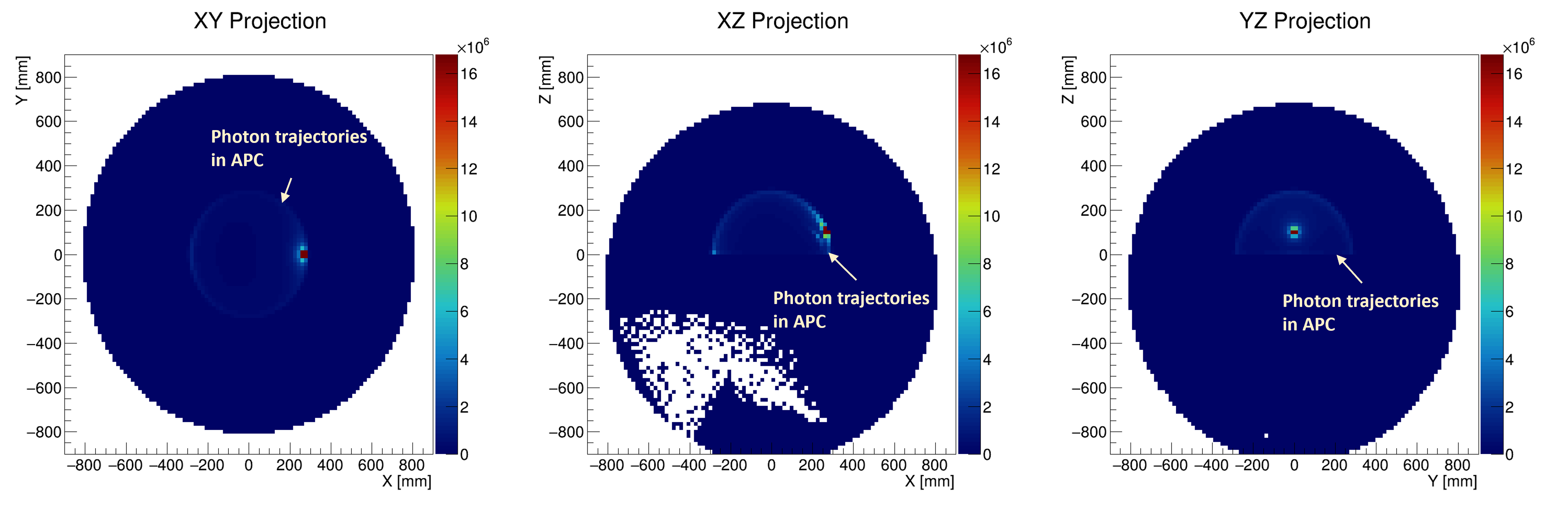}
    \caption{Results after reconstruction of photon traces}
    \label{photon_reconstruction}
\end{figure}

The propagation paths of photons generated after $\beta$ particles entered the APC were first reconstructed. The results are shown in figure~\ref{photon_reconstruction}, where the three subfigures represent the projections of photon trajectories onto the XY, XZ, and YZ planes, respectively. The contour of the APC is clearly visible in all projections, which is attributed to the fact that most Cherenkov photons undergo total internal reflection at the acrylic–air interface, becoming trapped and undergoing multiple reflections within the acrylic. Additionally, the photon production region is clearly identifiable and is primarily concentrated in a specific area on the outer layer of the acrylic, corresponding to the region where $\beta$ particles impinge on the material. In the XZ-plane projection, the blank region observed at the lower left of the APC arises from the obstruction of photon trajectories by the geometric structure of the PMTs.

\begin{figure}
    \centering
    \includegraphics[width=0.9\linewidth]{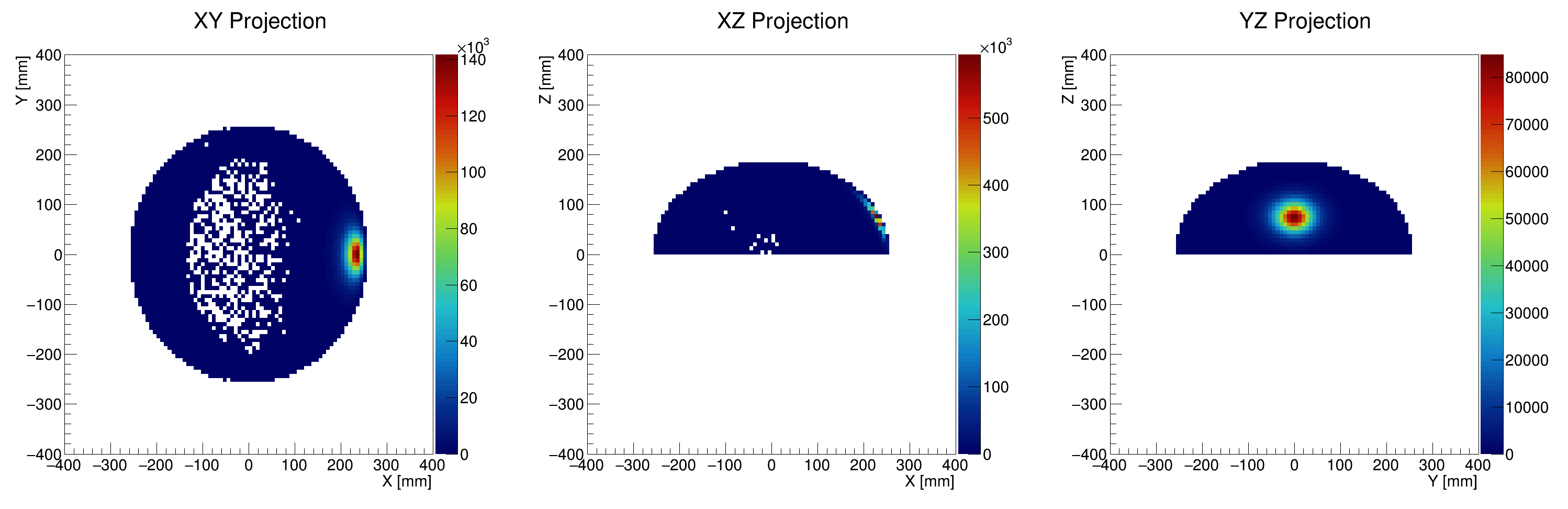}
    \caption{The incidence positions of photons on the photocathode surface, with the three panels representing their projected distributions on the XY, XZ, and YZ planes, respectively.}
    \label{fig:cut_trace}
\end{figure}

Subsequently, a selection was applied to retain only the photons incident on the photocathode, and their trajectories were reconstructed. The reconstruction results are shown in figure~\ref{fig:cut_trace}. 
A distinct ring-like structure can be observed in the region where photons exit the acrylic and reach the photocathode glass surface, which is highly consistent with the characteristic behavior of Cherenkov radiation. Moreover, the ring does not extend to the equatorial boundary of the APC, further indicating that the observed increase in DCR is independent of the geometrical arrangement. A comparison between the simulated result and the average DCR measured in the radioactive source experiment is summarized in table~\ref{tab:Comparison of simulation and experimental results}.

\begin{table}[htbp]
  \centering
  \renewcommand\arraystretch{1.2}  
  \begin{tabular}{lccc}
    \Xhline{1.2pt}
    & \makecell{Number of\\emitted particles} & \makecell{Increased\\DCR (Hz)} & \makecell{Proportion (Increased DCR /\\ Number of emitted particles)} \\
    \hline
    \makecell{Experimental \\results (average)} & 27646   & 14200   & 0.51 \\
    Simulated results                            & 1000000 & 459597  & 0.46 \\
    \hline
    \makecell{Difference } & --      & --      & $9.8\%$ \\
    \Xhline{1.2pt}
  \end{tabular}
  \caption{Comparison of simulation and experimental results. In particular, the DCR in the simulation results takes into account the quantum efficiency, and the average value of the PMT detection efficiency used in the JUNO experiment is 28\%\cite{abusleme2022mass}.}
  \label{tab:Comparison of simulation and experimental results}
\end{table}
In the experimental results, the number of emitted particles corresponds to the activity of the \(^{90}\mathrm{Sr}\) radioactive source, which is $2.7\times10^{4}~\mathrm{Bq}$.

The relative difference between the simulated results and the actual experimental measurements was determined to be $9.8\%$, indicating a strong agreement between the two results. This level of consistency validates the accuracy and reliability of the developed Geant4 model in replicating the physical processes of interest. Since the $\beta$ particles emitted by \(^{90}\mathrm{Sr}\) cannot penetrate the $10~\mathrm{mm}$ APC, this indicates that the observed increase in DCR is primarily attributable to Cherenkov light generated by the $\beta$ particles.

The present Geant4 model was designed to test the $\beta$-induced Cherenkov hypothesis with a localized \(^{90}\mathrm{Sr}\) source. Extending the simulation to the entire JUNO experimental hall is technically feasible and is part of our future programme. Such a study would require precise $\mathrm{U/Th/K}$ concentrations for the surrounding rock and concrete. By extending the simulation to compare the underground and surface environments, we can refine the DCR prediction.

\section{Conclusion}
In this experiment, we initially performed underground tests on the 20-inch MCP-PMTs. The test results indicate that when the PMT is placed in an underground environment similar to that of the JUNO detector, its DCR increases significantly, suggesting that the background radiation generated by the natural radioactivity of the surrounding rock may contribute to the increase in DCR. To verify this hypothesis, we constructed a lead shielding room in a laboratory located at the surface of JUNO experiment and tested the PMTs inside it. The results showed that the DCR of the PMTs decreased markedly when placed within the lead shield, confirming that lead effectively shields against environmental background radiation.

Subsequently, we irradiated the 20-inch MCP-PMTs with three different radioactive sources. We did not consider the effect of $\alpha$ particles on the PMTs due to the inherent very short penetration characteristics of $\alpha$ particles. The experiments revealed that both the X-rays produced by the \(^{55}\mathrm{Fe}\) source and the $\gamma$ rays emitted by the \(^{60}\mathrm{Co}\) source had negligible impact on the PMTs’ DCR, indicating that in the absence of a scintillator, neither $\mathrm{X}$ rays nor $\gamma$ rays interact significantly with the PMTs. In contrast, when a PMT was irradiated with $\beta$ particles from the \(^{90}\mathrm{Sr}\) source, its DCR increased significantly by approximately $14~\mathrm{kHz}$, demonstrating that $\beta$ particles can notably affect the PMT’s performance. 

Furthermore, the experimental results demonstrate that lead shielding can significantly reduce the DCR of the PMT, indicating that shielding materials such as water can be effectively employed in underground experiments to block environmental radioactive backgrounds and thereby suppress the DCR. This is of great significance for mitigating the impact of radioactive interference in underground experiments. Simultaneously, Geant4 simulations revealed that $\beta$ particles produce Cherenkov photons at the surface of the APC, and a fraction of this emission lies within the PMT photocathode’s response range. These photons reach the photocathode, generate photoelectrons through the photoelectric effect, and consequently increase the DCR. The simulation results are in good agreement with the experimental data, further supporting the indication that $\beta$ induced Cherenkov radiation is a major contributor to the observed DCR increase. However, a comprehensive simulation considering realistic JUNO site conditions would be required to fully quantify and confirm this effect.

\appendix



\bibliographystyle{JHEP}
\bibliography{biblio}

\end{document}